\documentclass[11pt]{amsart}
\usepackage{verbatim}
\usepackage[usenames]{color}
\usepackage{hyperref}

\newtheorem{theorem}{Theorem}
\newtheorem{proposition}[theorem]{Proposition}
\theoremstyle{remark}
\newtheorem{remark}{Remark}

\newcommand{\noprint}[1]{\relax}

\newcommand{\s}{\text{Sink}}

\title{Modular difference logic is hard}

\author[Bj{\o}rner]{Nikolaj Bj{\o}rner}
\address{Microsoft Research\\
  One Microsoft Way\\
  Redmond, WA \ 98052, USA} \email{nbjorner@microsoft.com}
\author[Blass]{Andreas Blass}
\address{Math Dept.\\
  University of Michigan\\
  Ann Arbor, MI 48109, USA} \email{ablass@umich.edu} \thanks{Partially
  supported by NSF grant DMS-0653696} \author[Gurevich]{Yuri Gurevich}
\address{Microsoft Research\\
  One Microsoft Way\\
  Redmond, WA \ 98052, USA} \email{gurevich@microsoft.com}
\author[Musuvathi]{Madan~Musuvathi}
\address{Microsoft Research\\
  One Microsoft Way\\
  Redmond, WA \ 98052, USA} \email{madanm@microsoft.com}
\begin{document}

\maketitle

\begin{abstract}
In connection with machine arithmetic, we are interested in
systems of constraints of the form $x + k \leq y + k'$.
Over integers, the satisfiability problem for such systems
is polynomial time. The problem becomes NP complete if we
restrict attention to the residues for a fixed modulus $N$.
\end{abstract}

\section{Introduction}\label{sec:intro}

The goal of this paper is to attract attention to the following problem: Given a system $\Sigma$ of inequalities, find out whether $\Sigma$ is satisfiable in a given machine arithmetic. We formalize a special case of the problem, in Section~\ref{sec:MDL}, as the satisfiability problem for modular difference logic (MDL). MDL is a variant of integer difference logic (IDL) described in Section~\ref{sec:IDL}. The IDL satisfiability problem admits a simple and efficient decision procedure. It turns out that the MDL satisfiability problem is infeasible (unless P$=$NP).

The MDL satisfiability problem is of particular relevance in the context of program verification and analysis. Established program verification environments~\cite{ESCModula} and abstract interpretation methods~\cite{Mine} have long relied on arithmetic over integers or over real numbers for reasoning about programs, and for a good reason. There are well known efficient methods for solving the satisfiability of linear arithmetic constraints over the reals, such as dual simplex or interior point methods. And integer constraints can be approached by extending simplex with Gomory cuts and branching methods; besides, important special cases, such as integer difference logic, admit efficient procedures. So the use of integer or even real semantics is well justified from the perspective of state of the art algorithms.

The obvious problem of course is that neither reals nor integers capture the proper semantics of machine arithmetic. Modular arithmetic, on the other hand, does capture machine arithmetic. Further, a system of difference constraints can be satisfiable over any fixed modulus $N>1$ but unsatisfiable over integers or reals, e.g.  $0 \leq x$ and $x + 1 \leq 0$. And a system of difference constraints can be satisfiable over integers and over reals, but unsatisfiable over a given modulus $N$, e.g. $x_0 < x_1 < \cdots < x_N$.

It follows that the common program analysis tools tend to rely on  methods that are both unsound and incomplete with respect to the accurate program semantics. We prove here that the MDL satisfiability problem is NP hard and thus the development of efficient tools for the MDL satisfiability problem is likely to be elusive. We also show that the problem is NP. The search for efficient methods for the MDL satisfiability problem is on.

\section{Integer difference logic}
\label{sec:IDL}

Integer difference logic (IDL) is a fragment of linear arithmetic.
IDL constraints have the form
\begin{equation*}
 x - y \ \leq\  k
\end{equation*}
where $x,y$ are integer variables and $k$ is an integer constant.
A system of IDL constraints may or may not have a solution.
For example, the constraint system
\begin{eqnarray*}
x_1 - x_2 \leq -3,\quad  x_2  - x_3 \leq 1,\quad x_3  - x_4 \leq -2,\quad x_4  - x_1 \leq 3
\end{eqnarray*}
is unsatisfiable, which can be established by adding the left and right-hand sides separately:
\begin{eqnarray*}
0 & = & (x_1  - x_2)+(x_2  - x_3)+(x_3  - x_4)+(x_4  - x_1) \\
 & \leq & -3 + 1 - 2 + 3 \\
 & = & -1
\end{eqnarray*}
The IDL constraint satisfiability problem (IDL-SAT) admits an
efficient decision procedure.

\begin{proposition}[\cite{Pratt,CM}]
IDL-SAT is solvable in polynomial time.
\end{proposition}

Some efficient procedures for IDL-SAT are  based on the Floyd-Warshall or Ford-Fulkerson style algorithms~\cite{Floyd,Warshall,FF}. IDL-SAT can be generalized to \emph{octagon} constraints $\pm x \pm y \leq k$ while still retaining polynomial time solvability~\cite{Mine}.

For the reader's convenience, we prove here the proposition. Our proof is based on the Floyd-Warshall algorithm.

\begin{proof}
Let $\Sigma$ be a system of IDL constraints. Without loss of generality, we presume that, for every pair $(x,y)$ of variables there is at most one constraint of the form $x-y \leq k$. Extend $\Sigma$ with an additional variable $\s$ adding constraints $x \leq \s$ (that is $x-\s \leq 0$) for every original variable $x$; given any solution for $\Sigma$, set $\s$ to the maximal value of the original variables to get a solution of the extended system $\Sigma^+$.

We construct a weighted directed graph $G$ on the variables of $\Sigma^+$: every constraint $x-y \leq k$ gives rise to an edge from $x$ to $y$ of weight $k$. In particular we have a weight-zero edge from any original variable $x$ to $\s$. If $G$ has a cycle of negative weight $-n$ that starts and ends at vertex $x$ then an unsolvable constraint $x - x \leq -n $ is obtained by adding the inequalities from $\Sigma^+$ that gave rise to the edges in the cycle.

The polynomial-time Floyd-Warshall algorithm~\cite{Floyd,Warshall} finds out whether $G$ has negative cycles. Furthermore, suppose that $G$ has no negative cycles. Then the Floyd-Warshall algorithm computes the minimal weight $W(x,y)$ of any path from $x$ to $y$; if there is no path from $x$ to $y$ then $W(x,y) = \infty$. This allows us to construct a solution $S$ for $\Sigma$.

Set $S(\s) = 0$ and $S(x) = W(x,\s)$ for every variable in $\Sigma$. Every constraint $x-y \leq k$ of $\Sigma$ is satisfied. Indeed, by the minimality of $W$, we have $W(x,\s) \leq W(x,y) + W(y,\s)$ and $W(x,y) \leq k$. Hence $S(x) \leq k + S(y)$ and $S(x) - S(y) \leq k$. (Note that $W(x,\s)$ is the minimal weight of any path from $x$ on the original variables, so $\s$ is not really needed.)
\end{proof}

But integer difference logic cannot be directly used when reasoning about constraints coming from machine arithmetic because machine arithmetic uses modular addition.  The question thus arises what is the complexity of the constraint satisfiability problem in the case of modular arithmetic? We establish here that the problem is NP complete.

\section{Modular difference logic}
\label{sec:MDL}

Modular difference logic (MDL) is similar to integer difference logic
except that integers are replaced with residues $0, 1, \ldots, N-1$
modulo a fixed positive integer $N$. The residues are ordered in the
obvious way; the maximal residue is $N-1$.

Instead of restricting attention to the residues, it may be beneficial
to work, modulo $N$, with arbitrary integers, and we will often do
that. But one should be careful not to confuse (a)~the standard
integer order $\leq$ and (b)~another relation on integers, which we
call $\leq_N$ and will define shortly, that reflects the order of the
residues. Each integer $i$ is equal modulo $N$ to a unique residue
$i_N$. Define $i\leq_N j$ if $i_N \leq j_N$. Relations $=_N, \geq_N,
<_N, >_N$ are defined accordingly. These definitions precisely match
the semantics of comparison operations supported by current hardware
architectures for machine arithmetic.

In the case of integers, a constraint $x - y \leq k$ is equivalent to
constraint $x \leq y + k$. This is not necessarily true in modular
arithmetic.
For example $9 - 5 \leq_{10} 5$ but $9 \not\leq_{10} 5+5$.
Similarly $x + 1 \leq_N y$ is not necessarily equivalent to $x \leq_N
y - 1$. For example, $5 \leq_{10} 0 - 1 =_{10} 9$ but $5+1 >_{10} 0$.

We define MDL constraints to have the form
\begin{eqnarray}\label{eq:form0}
x + k \ \leq_N\  y + \ell
\end{eqnarray}
where $x,y$ are variables and $k,\ell$ are constants.  The MDL
Satisfiability Problem (MDL-SAT) is the satisfiability problem for
systems of MDL constraints.

\begin{remark}\label{rem:logic}
From the point of view of logic, modular difference logic is a
fragment of the first-order theory $T$ of discrete linear order with
both ends (and two constants for the two ends) and with the cyclic
successor and predecessor function. The two constants could be called
Min and Max. The successor of Max is Min, and the predecessor of Min
is Max.  The question arises what's $x+k$? This depends on the sign of
$k$. If $k\geq 0$ then $x+k$ is the result of $k$-fold application of
the successor function to $x$; otherwise $x+k$ is the result of
$|k|$-fold application of the predecessor function. The residues
modulo $N$ form a model of $T$ where Min = $0$ and Max = $N-1$. There
are also infinite models of $T$. One of them can be obtained by
reordering the integers as follows:
\[
 0 < 1 < 2 < 3 < \cdots < -3 < -2 < -1.
\]
This order is reminiscent of the order $\leq_N$, where $-1$ is also the
maximal element.

It is known (and not hard to check, by means of an
Ehrenfeucht-Fra{\"\i}ss{\'e} game \cite{EF}) that,
for every first-order sentence $\phi$ in
the language of $T$, there is a natural number $n$, such that $\phi$
does not distinguish between any two models of $T$ of size $\geq
n$. It follows that all infinite models of $T$ are elementarily
equivalent.  In that sense, one may speak about the infinite model of
$T$.

We are interested primarily in the case of a modulus $N$ that is
large. From the point of view of logic, we can as well work with the
infinite model of $T$. Every constraint-satisfaction problem for MDL
can be formulated as an existential sentence in the language of
$T$.\qed
\end{remark}

\section{MDL-SAT is NP hard}
\label{sec:MDL-is-hard}

We now establish that a very modest fragment of MDL-SAT is NP hard.

\begin{theorem}\label{th:1}
Suppose that $N \geq 4$. Then the fragment of MDL-SAT with constraints
of the form
\begin{equation}\label{eq:form1}
 x + 1 \ \leq_N\  y\quad \text{or}\quad x \ \leq_N\  y -1
\end{equation}
is NP hard.
\end{theorem}

\begin{proof}
Given a graph $G$, we construct a system of MDL constraints that is
satisfiable if and only if the graph is 3-colorable. It will be
convenient to assume that the vertices of $G$ are linearly
ordered. This allows us to represent edges as ordered pairs $(v,w)$
where $v<w$.

With every vertex $v$ of $G$ we associate three variables $v_0, v_1,$
and $v_2$ and three constraints
\begin{equation}\label{eq:1a}
\begin{split}
 v_0 + 1 \ &\leq_N\ v_1\\
 v_1 + 1 \ &\leq_N\ v_2\\
 v_2 + 1 \ &\leq_N\ v_0.
\end{split}
\end{equation}

One consequence of constraints~\eqref{eq:1a} is that at least one of the three variables takes the maximal value $N-1$. With each edge $e = (v,w)$ we associate six variables $e_1, e_2, e_3, f_1, f_2, f_3$ and nine constraints: three constraints
\begin{equation}\label{eq:1b}
\begin{split}
	v_c      &\ \leq_N\  e_c - 1,\\
 	w_c      &\ \leq_N\  f_c -1, \\
 	f_c + 1  &\ \leq_N\  e_c
\end{split}
\end{equation}
for each $c = 0, 1, 2$. One consequence of the three constraints~\eqref{eq:1b} is that residues $v_c$ and $w_c$ cannot simultaneously have the maximal value $N-1$. Indeed, if $v_c = w_c = N-1$ then, by the first and second constraints, $e_c = f_c = 0$ which contradicts the third constraint. If all the constraints are satisfied then we have a 3-coloring for $G$: the color of a vertex $v$ is the first number $c$ such that $v_c = N-1$. By \eqref{eq:1a}, every vertex has a unique color. By \eqref{eq:1b}, no two adjacent vertices have the same color.

Now we suppose that $G$ is 3-colorable (with colors $0,1,2$) and prove that the constraint system is satisfiable. For every color $c$ and every vertex $v$ of color $c$, set
\[
 v_c = N-1,\quad v_{c+1} = 0,\quad v_{c+2} = 1.
\]
where addition in the subscripts is modulo 3. Clearly all inequalities~\eqref{eq:1a} are satisfied. Now consider an edge $e = (v,w)$ and a color $c$. We show how to satisfy the three constraints~\eqref{eq:1b}.

Case 1: $c$ is the color of $v$, so that $v_c = N-1$. Since $w$ does not have color $c$, we have $w_c \in \{0,1\}$. To satisfy the first of the three constraints, set $e_c = 0$. To satisfy the third constraint, set $f_c = N - 1$. The second constraint is satisfied as well: $w_c \leq 1 \leq N-2$.

Case 2: $c$ is the color of $w$, so that $w_c = N-1$ and $v_c \in \{0,1\}$. To satisfy the second constraint, set $f_c = 0$. To satisfy the first and third constraints, set $e_c = 2$.

Case 3: neither $v$ nor $w$ is of color $c$, so that both $v_c$ and $w_c$ are $\leq 1$. Set $f_c = 2$ and $e_c = 3$.
\end{proof}

\begin{remark}
One may be interested in the variant of MDL-SAT where the modulus $N$ is not fixed but is a part of the input. Theorem~\ref{th:1} and its proof remain valid.
\end{remark}

\section{Strict Inequalities}

Over integers, a non-strict inequality $x - y \leq k$ is equivalent to
a strict inequality $x - y < k+1$. The relation between non-strict and
strict inequalities is much more subtle in modular arithmetic. With
this in mind, we prove a version of Theorem~\ref{th:1} with strict
inequalities.

\begin{theorem}\label{th:2}
Suppose that $N\geq 9$. Then the fragment of the modified MDL-SAT with
constraints of the form
\begin{equation*}
 x + k <_N y + \ell
\end{equation*}
is NP hard.
\end{theorem}

In fact, we will use only values $0,1,2$ for $k$ and only values
$0,1,-1$ for $\ell$.

\begin{proof}
The proof is again by reduction from the 3-colorability problem, and
it is similar to the proof of
Theorem~\ref{th:1}. Constraints~\eqref{eq:1a} replaced with constraints

\begin{equation}\label{eq:2a}
\begin{split}
 v_0 + 2 \ &<_N\ v_1,\\
 v_1 + 2 \ &<_N\ v_2,\\
 v_2 + 2 \ &<_N\ v_0,
\end{split}
\end{equation}
and constraints~\eqref{eq:1b} are replaced with constraints
\begin{equation}\label{eq:2b}
\begin{split}
	v_c      &\ <_N\  e_c - 1,\\
 	w_c      &\ <_N\  f_c - 1, \\
 	f_c + 1  &\ <_N\  e_c + 1
\end{split}
\end{equation}

For each vertex $v$, constraints~\eqref{eq:2a} force at least one of the
three residues $v_c$ to be $\geq N-2$.
The idea is that when $v_c$
has value $\geq N-2$, then $c$ is an acceptable color for $v$.
Constraints~\eqref{eq:2b} imply that residues $v_c$ and $w_c$ cannot be
simultaneously $\geq N-2$. Indeed, by the first of the three
constraints, $v_c$ cannot have the maximal value $N-1$, and if $v_c =
N-2$ then $e_c = 0$.  Similarly, $w_c \neq N-1$, and if $w_c = N-2$
then $f_c = 0$. If $v_c = w_c = N-2$ then $e_c = f_c = 0$ and then the
third inequality fails.
Thus, any solution of the new system of constraints yields a
3-coloring of $G$.

In the other direction, we need to convert a given 3-coloring of $G$
into a solution for the constraint system. For every color $c$ and
every vertex $v$ of color $c$, we set
\begin{eqnarray*}
    v_c = N-2,  v_{c+1} = 1,  v_{c+2} = 4.
\end{eqnarray*}
Clearly \eqref{eq:2a} is satisfied.  Now consider an edge $e = (v,w)$
and a color $c$. We show how to satisfy the three constraints
\eqref{eq:2b}. As in the proof of Theorem~\ref{th:1}, we consider three
cases.

Case 1: $c$ is the color of $v$, so that $v_c = N-2$ and $w_c \in
\{1,4\}$. To satisfy the first constraint, set $e_c = 0$.  To satisfy
the third constraint, set $f_c = N-1$.  The second constraint is
satisfied as $w_c \leq 4 < N - 2 = f_c - 1$.

Case 2: $c$ is the color of $w$ so that $w_c = N-2$ and $v_c \in
\{1,4\}$. Set $e_c = 6$ and $f_c = 0$.  Clearly \eqref{eq:2b} is
satisfied.

Case 3: Neither $v$ nor $w$ is of color $c$ so that both $v_c$ and
$w_c$ are in $\{1,4\}$.  Set $e_c = 7$ and $f_c = 6$.
\end{proof}

\section{MDL-SAT is NP}

In this section, we modify the satisfiability problem MDL-SAT for
modular difference logic in two ways.
First, the modulus $N$ is a part
of the input. Second, we liberalize the notion of MDL constraints by
allowing constraints in the form of non-strict inequalities of the
form
\[
 x + k \ \leq_N\  y + \ell, \quad\text{or}\quad x \ \leq_N\ k, \quad \text{or}\quad x \ \geq_N\ k,
\]
strict inequalities of the form
\[
 x + k \ <_N\ y + \ell, \quad\text{or}\quad x \ <_N\ k, \quad \text{or}\quad x \ >_N\ k,
\]
as well as equalities of the form
\[
 x + k \ =_N\ y + \ell, \quad\text{or}\quad x \ =_N\ k.
\]
Both modifications make the problem harder and thus make the next theorem stronger.

\begin{theorem}\label{th:3}
The constraint satisfiability problem MDL-SAT for modular difference logic is  NP.
\end{theorem}

\begin{proof}
Let $\Sigma$ be a system of MDL constraints with $p$ variables. Let $m$ be the maximum of the absolute values of the constants in the $\Sigma$ constraints. We prove that, if $\Sigma$ has any solution, then it has a solution where the absolute values of all variables are $\leq (2m+1)p$. It follows that MDL-SAT is NP.

Suppose that $\Sigma$ has a solution $S$ that maps the variables into the residues modulo $N$. To simplify the exposition, we extend $\Sigma$ with two additional variables $v_{\min}, v_{\max}$ and with two equations $v_{\min} = 0$, $v_{\max} = -1$. The solution $S$ extends appropriately.

We create an auxiliary graph $G_S$. The vertices are the variables of $\Sigma$, and the edges are pairs $\{v,w\}$ such that $|S(v) - S(w)| \leq 2m$. Connected components of $G_S$ will be called \emph{clusters}. The \emph{domain} of a cluster $C$ is a closed interval $[a,b]$. If $v$ is a leftmost variable of $C$ (so that $S(v) \leq S(w)$ for any other variable $w\in C$) then $a = \max\{0,S(v) - m\}$. And if $v$ is the rightmost variable of $C$ then $b = \min\{N-1,S(v) + m\}$. The domains of different clusters are disjoint.

The clusters different from those of $v_{\min}$ and $v_{\max}$ will be called \emph{inner}. The crucial observation is that inner clusters could be shifted around. Indeed, consider an inner cluster $C$ with domain $[a,b]$, and let $r$ be the right end of the domain of the left neighbor of $C$, so that $a>r$. If $r < a' < a$, shift $C$ leftward for distance $d = a - a'$, that is, modify assignment $S$ to an assignment $S'$ that is like $S$ except that $S'(v) = S(v) - d$ on the variables $v$ of $C$. It is easy to see that $S'$ is a solution for $\Sigma$. In a similar way clusters could be shifted to the right.

Now we are ready to produce the desired small-value solution. If there are inner clusters, shift the leftmost inner cluster $C_1$ to the left as far as possible (so that $a' = r+1$ in the notation of the previous paragraph). If there are inner clusters to the right of $C_1$, shift the right neighbor $C_2$ of $C_1$ to the left as far as possible. And so on until all inner clusters are packed as close as possible on the left side. Let $S^*$ be the resulting solution. In the rest of the proof, variables represent their $S^*$ values.

In addition to $v_{\min} = 0$, there are $\ell \leq p$ original variables in the cluster of $v_{\min}$ and the inner clusters: $v_0 = v_{\min} < v_1 < \cdots < v_\ell$. Every $v_{i+1} - v_i \leq 2m + 1$. It follows that every $v_i \leq v_\ell \leq \ell(2m+1) \leq p(2m+1)$. A similar argument applies to the cluster of $v_{\max}$ except that there the distance between neighboring variables is $\leq 2m$. Every variable $v$ there is within distance $2pm$ from the end, so that $|v| \leq 2pm+1$. That completes the proof.
\end{proof}

\begin{remark}
We have not used the fact that modulus $N$ is a part of the input. The theorem and the proof remain valid if the modulus is fixed or even if it is infinite as in Remark~\ref{rem:logic}.
\end{remark}

\bibliographystyle{plain}
\bibliography{refs}

\end{document}